\documentclass[pre,a4paper,superscriptaddress,showpacs]{revtex4}
\usepackage{graphicx}
\begin{document}

\newcommand{\EQ}{Eq.~}
\newcommand{\EQS}{Eqs.~}
\newcommand{\FIG}{Fig.~}
\newcommand{\FIGS}{Figs.~}
\newcommand{\TAB}{Tab.~}
\newcommand{\SEC}{Sec.~}
\newcommand{\SECS}{Secs.~}

\title{Return times of random walk on generalized random graphs}
\author{Naoki Masuda}
\affiliation{Laboratory for Mathematical Neuroscience,
RIKEN Brain Science Institute, 2-1, Hirosawa, Wako, Saitama, 351-0198,
Japan}
\author{Norio Konno}
\affiliation{Faculty of Engineering,
Yokohama National University,
79-5, Tokiwadai, Hodogaya, Yokohama, 240-8501 Japan}
%
\date{Received 14 November 2003}

\begin{abstract}
Random walks are used for modeling various dynamics in, for
example, physical, biological, and social contexts.  Furthermore,
their characteristics provide us with useful information on the phase
transition and critical phenomena of even broader classes of related
stochastic models. Abundant results are obtained for random walk on
simple graphs such as the regular lattices and the Cayley
trees. However, random walks and related processes on more complex
networks, which are often more relevant in the real world, are still
open issues, possibly yielding different characteristics.  In this
paper, we investigate the return times of random walks on random
graphs with arbitrary vertex degree distributions. We analytically
derive the distributions of the return times.  The
results are applied to some types of networks and compared with
numerical data.
\end{abstract}


\maketitle

\section{Introduction}\label{sec:introduction}

The theory of the random walk has a long history.  Random walks and their
extensions
have also been applied
 with profound theoretical bases to modeling numerous types of
physical, biological, sociological, and economical dynamics
\cite{Spitzer}. For example, distributions of return times
and scaled limit distributions of the walkers' positions
are broadly known for simple underlying graphs. They provide
useful information on critical values and phase transitions
in regard to survival of the branching random walks and the contact
processes \cite{Liggettbook99,Schinazibook,Pemantle01}, survival in the voter
models \cite{Liggettbook99,Durrettbook}, and
occurrence of percolation
\cite{Lyons90_Lyons95},

Indeed, a large body of theoretical results are available for random
walks performed on regular lattices such as ${\bf Z}^d$ and on the
Cayley (or regular) trees, which are defined to be trees with
homogeneous vertex degree. However, it has been suggested recently
that more complex networks as opposed to regular graphs and
conventional random graphs \cite{Erdos} are concerned to real worlds.
Particularly, important classes of random graphs such as small-world
networks and scale-free networks were proposed and have been examined
in the last several years. These networks share some important
properties with real networks, such as the clustering property, short
average path length, and the power-law of the vertex degree
distributions \cite{SW-SF,Callaway00,Albert02,Newman}.  They have been
applied to the analysis of various biological, engineering, and social
networks including information flow in the Internet
\cite{Callaway00,Albert02,Newman} and epidemics
\cite{Newman,Eguiluz_Pastor_Masuda_SARS}. The
properties of spatial stochastic models, both static configurations
and dynamical processes, typically change as the network topology
varies even when other basic quantities such as the mean vertex degree
is conserved. For example, the analysis of percolation-based models
revealed that the critical parameter values for the
occurrence of global epidemics, or even their existence, depend on
network topology \cite{Callaway00,Albert02,Newman}.

It is highly likely that the properties of random walks depend on
network topology \cite{Schinazibook,Pemantle01}, as numerical and
approximate results suggest for the quenched \cite{Pandit}
and annealed \cite{Jasch_Lahtinen} Watts-Strogatz-type small-world
networks and for quenched random graphs with homogeneous vertex
degree \cite{Szabo00rw}. In relation to this issue, how eigenvalues of
the adjacency matrices are distributed has been numerically examined
for scale-free and small-world networks \cite{Farkas_Goh}. The largest
eigenvalue of an adjacency matrix measures how the number of closed
paths increases as the path length tends to
infinity.  The eigenvalues supply useful information on the return
times of random walks \cite{Spitzer}, serving to a wide range of
applications as mentioned above.  However, the largest eigenvalue
$\rho$ has
been characterized only in terms of the numerical scaling law
for the scale-free networks 
in an unnormalized manner, namely, $\rho\propto
m^{1/2}N^{1/4}$, where $N$ is the system size and $2m$ is the mean
vertex degree \cite{Farkas_Goh}.

In this paper, we analyze random walks on a general class of random
graphs that includes random scale-free networks, the
Erd\"{o}s-R\'{e}nyi random graph, and the Cayley trees as special
cases \cite{Lyons90_Lyons95,Pemantle01,Albert02,Newman}.  Explicit
expressions for the first return time probability and the annealed
approximation forms for the general
return time probability are derived with the use of partition of
integers. In \SEC\ref{sec:model}, we introduce the network model and
the generating functions.  In
\SEC\ref{sec:return}, we calculate the probability distribution
functions of the return time of random walk. Then, in
\SEC\ref{sec:examples}, we confirm with some
examples that our theoretical estimates match
numerical results. Lastly, the conclusion
follows, and the difference in the decay
rate of the return time probability between regular and random
networks,
which implies the difference in
the possibility of percolation and the survival of contact processes,
are also touched upon.
 
\section{Network Model and Generating Functions}\label{sec:model}

We analyze a class of random graphs called generalized random graphs
in physical contexts \cite{Callaway00,Albert02,Newman} or
Galton-Watson trees in mathematical contexts
\cite{Lyons90_Lyons95,Pemantle01}.  These random graphs are infinite
trees without loops. The degree of each vertex, or the number of
neighbors, is distributed according to an identical and independent
probability density function.  As shown in \FIG\ref{fig:rw_pic},
each realization of the graph, which is generally inhomogeneous,
is taken from the random ensemble. However,
they are regular in a
statistical sense.  Let us denote by $p_k$ the probability that a
vertex has the degree equal to $k$. We assume that $p_0=0$ without
losing generality.  Consequently, $\sum^{\infty}_{k=1}p_k=1$.

Let us designate an arbitrary vertex $O$ of a realized graph
as the root. We examine a random walk
starting from $O$. Since we exclusively deal with trees here, the
random walker can return to $O$ only when the time $n\in$ $\{0, 1, 2,
\ldots \}$ is even. In accordance, we denote by $q_n$ the probability
that the random walker returns to $O$ for the first time at time $2n$,
and by $r_n$ the probability that it returns to $O$ irrespective of
the accumulated number of returns. Here we consider only the
annealed random walk, confining ourselves in the analysis of return
times averaged over both probability space of graph and that of
random walk. To be contrasted with the annealed randomness 
is the quenched randomness, which
is concerned to the
ensemble of walkers on a fixed realization of random graph
\cite{Zeitouni02}. Both quenched
\cite{Pandit,Szabo00rw} and annealed \cite{Jasch_Lahtinen} random walks have
been implicitly treated in the studies of
random dynamics on complex networks.
Though quenched environments are realistic,
the statistics based on annealed walks that we derive in the
following
can be regarded as averages of the statistics of quenched walks
over the ensemble of a random graph.

The generating functions for the distributions 
$\{q_n\}$ and $\{r_n\}$, which we respectively denote by
$Q(z)$ and $R(z)$ are defined by
\begin{equation}
Q(z)\equiv \sum^{\infty}_{n=0} q_n z^n,\quad
R(z)\equiv \sum^{\infty}_{n=0} r_n z^n.
\end{equation}
With $q_0 = 0$ and $r_0 = 1$, $Q(z)$ and $R(z)$ satisfy the following
relation:
\begin{eqnarray}
R(z) &=& \sum^{\infty}_{n=0} 
\left( \sum^{n}_{m=1} q_m r_{n-m} + \delta_{n,0}\right) z^n\nonumber\\
&=& R(z)Q(z) + 1,
\label{eq:Q_R}
\end{eqnarray}
where $\delta_{i,j} = 1$ for $i=j$ and $\delta_{i,j} = 0$
otherwise \cite{Spitzer,Grimmettbook}.
Strictly speaking, \EQ(\ref{eq:Q_R}) is valid only for the quenched
case. Therefore, the following results for $R(z)$ should be understood
as an approximation by annealed statistics.

\section{Distributions of Return Times}\label{sec:return}

\begin{figure}[b]
\begin{center}
\includegraphics[height=2.25in,width=3.25in]{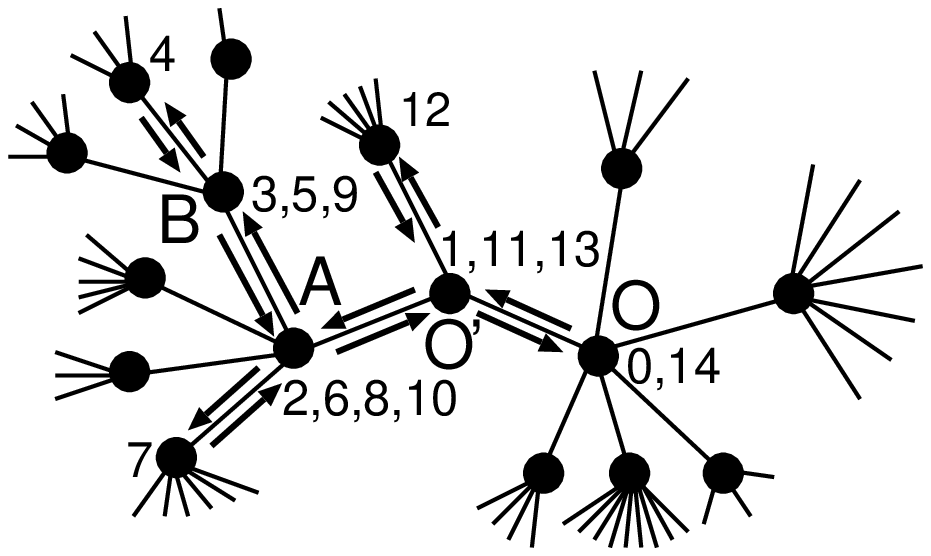}
\caption{}
\label{fig:rw_pic}
\end{center}
\end{figure}

To derive explicit expressions for the return time
distributions, we provide the
approximate recursion relation below.
Let us resort to \FIG\ref{fig:rw_pic} for explanation.
Suppose that the random walker starting from $O$ returns to $O$ after $2n$
steps for the first time ($n=7$ in \FIG\ref{fig:rw_pic}).
In the first step, the random walker moves
to a neighbor that we denote by $O^{\prime}$. Because of the
statistical homogeneity of the generalized random graph, the vertex
degree of $O^{\prime}$ is distributed as specified by $\{ p_k\}$
whichever neighbor of $O$ is chosen.  The random walker has to arrive
at $O^{\prime}$ at time $2n-1$ ($=13$) and move to $O$ subsequently at time
$2n$ $(=14)$.
The last event occurs with probability $1/k$.  In the meantime,
the random walker travels for $2n-2$ steps without visiting
$O$. The walker wanders in the subtrees rooted at
$O^{\prime}$ to complete loops, or closed paths of random walk.
Any such loop cannot contain $O$, and the probability that a
path emanating from $O^{\prime}$ enters a subtree is $(k-1)/k$.  Let us
denote by $a$ the number of the loops originating from
$O^{\prime}$. In \FIG\ref{fig:rw_pic}, $a$ is equal to 2. 
Then, the length $2n_i$ ($1\le i\le a$) of each loop is
even ($n_1=5$ and $n_2=1$ in \FIG\ref{fig:rw_pic}),
and $2n_i$ must sum up to $2n-2$. In addition, since the vertex
degree is homogeneously distributed, the probability law for 
the length of loop is assumed to be the same as that for
the original random walk starting and ending at $O$.

Here we make a crucial approximation of disregarding any
memory effects. In other words, we suppose that the $a$ subtrees
rooted at $O^{\prime}$ are
independent of each other. In fact, if the same neighbor of
$O^{\prime}$ is chosen for different entries into the subtree, the
subtrees reached by these different 
entries coincide. As an example, the random walker shown in
\FIG\ref{fig:rw_pic} travels from $A$ to
the subtree rooted at $B$ twice, before returning to
$O$. In
this occasion, it is not qualified to regard the vertex degrees and
the loop lengths to be independent for 
the two neighbors of $O^{\prime}$. However, the
approximation error is small unless the mean vertex
degree is extremely small. The accuracy of the 
following analytical methods are
investigated in comparison with
numerical simulations in \SEC\ref{sec:examples}. 

Based on the consideration above, we have the following recursion formula:
\begin{eqnarray}
q_n &=& \sum^{\infty}_{k=1}p_k \sum^{n-1}_{a=0}
\sum_{\sum^a_{i=1} n_i = n-1, n_i\ge 0, 1\le i\le a}
 \frac{k-1}{k}q_{n_1}\; \frac{k-1}{k}q_{n_2}\;
\ldots \; \frac{k-1}{k}q_{n_a} \; \frac{1}{k}\nonumber\\
&=& \sum^{\infty}_{k=1}\frac{p_k}{k} \sum^{\infty}_{a=0}
\left(\frac{k-1}{k}\right)^a
\sum_{\sum^a_{i=1} n_i = n-1, n_i\ge 0, 1\le i\le a} 
\prod^a_{a^{\prime}=1} q_{n_{a^{\prime}}},
\label{eq:q_n}
\end{eqnarray}
which covers the singular case $q_0=0$ as well.
Using \EQ(\ref{eq:q_n}), the generating function of $q_n$ is
calculated as
\begin{eqnarray}
Q(z) &=& \sum^{\infty}_{n=0} \left\{
\sum^{\infty}_{k=1}\frac{p_k}{k} \sum^{\infty}_{a=0}
\left(\frac{k-1}{k}\right)^a
\sum_{\sum^a_{i=1} n_i = n-1, n_i\ge 0, 1\le i\le a} 
\prod^a_{a^{\prime}=1} q_{n_{a^{\prime}}} \right\} z^n\nonumber\\
&=& z 
\sum^{\infty}_{k=1}\frac{p_k}{k} \sum^{\infty}_{a=0}
\left(\frac{k-1}{k}\right)^a
 \sum^{\infty}_{n=0}
\sum_{\sum^a_{i=1} n_i = n-1, n_i\ge 0, 1\le i\le a} 
\prod^a_{a^{\prime}=1} q_{n_{a^{\prime}}}z^{n_a^{\prime}} \nonumber\\
&=& 
z \sum^{\infty}_{k=1}\frac{p_k}{k} \sum^{\infty}_{a=0}
\left(\frac{k-1}{k}\right)^a
Q(z)^a\nonumber\\
&=& z \sum^{\infty}_{k=1}\frac{p_k}{k-(k-1)Q(z)}.
\label{eq:Q}
\end{eqnarray}
Although $Q(1)=1$ is always consistent with \EQ(\ref{eq:Q}),
we exclude this case because the random walk on generalized random
graphs including the Cayley trees
is transient \cite{Lyons90_Lyons95}, except for the Cayley tree
with vertex degree 2, which is identical to ${\bf Z}$.
Accordingly, we look for the solution satisfying $Q(1)<1$.

By expanding the right-hand side of
\EQ(\ref{eq:Q}), we obtain
\begin{eqnarray}
Q(z) &=& z \sum^{\infty}_{k=1} \frac{p_k}{k} 
\sum^{\infty}_{n=0} \left( \frac{k-1}{k}Q(z) \right)^n\nonumber\\
&=& \frac{z M\left[Q(z)\right]}{Q(z)},
\label{eq:Q_expand}
\end{eqnarray}
where
\begin{equation}
M(z)\equiv \sum^{\infty}_{n=1} m_n z^n
\end{equation}
is the generating function of the moment function
given by
\begin{equation}
m_n \equiv \sum^{\infty}_{k=1} \frac{(k-1)^{n-1}}{k^n}\; p_k.
\end{equation}
In deriving \EQ(\ref{eq:Q_expand}), the expansion is justified by the
fact that $Q(z)$ has the radius of convergence equal to 1 and that
$(k-1)/k<1$.  Then, we apply the following theorem to calculate
$Q(z)$ and $R(z)$.

{\it Lagrange's inversion formula} \cite{Grimmettbook}
Let $z=w/f(w)$ where $w/f(w)$ is an analytic function of $w$ near
$w=0$. If $g$ is infinitely differentiable, then
\begin{equation}
g\left(w(z)\right) = g(0) + \sum^{\infty}_{n=1}\frac{z^n}{n!}
\left[ \frac{d^{n-1}}{du^{n-1}} 
\left[ g^{\prime}(u)f(u)^n\right] \right]_{u=0}.
\label{eq:Lagrange}
\end{equation}

For our purpose, we set $w(z)=Q(z)$, $f(w) = M(w)/w$, $g(w)=w$ in
\EQ(\ref{eq:Lagrange}).  Apparently, the fact that
$m_1>0$ guarantees the regularity
of $w/f(w)$ around $w=0$.  As a result, we have
\begin{eqnarray}
Q(z) &=& g(w)\nonumber\\
&=& 0 + 
\sum^{\infty}_{n=1}\frac{z^n}{n!}
\left\{ \frac{d^{n-1}}{du^{n-1}} 
\left[ \left(\frac{M(u)}{u}\right)^n\right] \right\}_{u=0}.
\label{eq:Q_Lagrange}
\end{eqnarray}
We also note that
\begin{eqnarray}
M(u)^n &=& (m_1 u + m_2 u^2 + \cdots )^n\nonumber\\
&=& \sum^{\infty}_{n^{\prime}=n} u^{n^{\prime}}
\sum_{\lambda\vdash n^{\prime},
\sum^{\infty}_{l=1} i_{\lambda}(l) = n}
\frac{n!}{\prod^{\infty}_{l=1}i_{\lambda}(l)!}
\prod^{\infty}_{l=1} m_l^{i_{\lambda}(l)},
\label{eq:M_power_n}
\end{eqnarray}
where $\sum_{\lambda\vdash n^{\prime}}$ indicates 
the summation over all the partitions of the integer 
$n^{\prime}$ into integers. In general, a partition $\lambda$
is represented by
$\lambda=(1^{i_{\lambda}(1)} 2^{i_{\lambda}(2)} \cdots )$, which
means that 1 is included $i_{\lambda}(1)$
times in $\lambda$,
2 is included $i_{\lambda}(2)$ times, and so on \cite{Stanleybook}.
By the definition of partition, $\{i_{\lambda}(1), i_{\lambda}(2), \ldots
\}$ ($\lambda\vdash n^{\prime}$) 
satisfies 
\begin{equation}
\sum^{\infty}_{l=1} l\; i_{\lambda}(l) 
= \sum^{n^{\prime}}_{l=1} l\; i_{\lambda}(l) = n^{\prime}. 
\end{equation}
For example,
\begin{eqnarray}
\left\{ \lambda \; | \; \lambda\vdash 5 \right\} &=&
\left\{(1^5), (1^3 2), (1^2 3), (1 2^2), (1 4), (2 3), (5) \right\},
\label{eq:partition_ex_1}\\
\left\{ \lambda \; | \; \lambda\vdash 7 \right\} &=&
\left\{(1^7), (1^5 2), (1^4 3), (1^3 2^2), (1^3 4),
(1^2 2 3), (1^2 5), (1 2^3), (1 2 4), (1 3^2), \right.\nonumber\\
&& \left. (1 6), (2^2 3), (2 5),
(3 4), (7) \right\}.
\label{eq:partition_ex_2}
\end{eqnarray}
In \EQ(\ref{eq:M_power_n}), only the partitions whose numbers of parts
are $n$ are concerned.  Corresponding to \EQS(\ref{eq:partition_ex_1})
and (\ref{eq:partition_ex_2}), the partitions appearing in the
summation of \EQ(\ref{eq:M_power_n}) for $(n,n^{\prime})=(3,5)$ and
$(4,7)$ are as follows:
\begin{eqnarray}
\left\{ \lambda \; \bigg| \; \lambda\vdash 5, \sum^{\infty}_{l=1}
i_{\lambda}(l) = 3\right\} &=& \left\{ (1^2 3), (1 2^2) \right\},\\
\left\{ \lambda \; \bigg| \; \lambda\vdash 7, \sum^{\infty}_{l=1}
i_{\lambda}(l) = 4\right\} &=& 
\left\{ (1^3 4), (1^2 2 3), (1 2^3) \right\}.
\end{eqnarray}
With \EQ(\ref{eq:M_power_n}), \EQ(\ref{eq:Q_Lagrange}) is evaluated
as follows:
\begin{eqnarray}
Q(z) &=&
\sum^{\infty}_{n=1}\frac{z^n}{n!}
(n-1)!
\sum_{\lambda\vdash 2n-1,
\sum^{\infty}_{l=1} i_{\lambda}(l) = n}
\frac{n!}{\prod^{\infty}_{l=1}i_{\lambda}(l)!}
\prod^{\infty}_{l=1} m_l^{i_{\lambda}(l)}\nonumber\\
&=&
\sum^{\infty}_{n=1} z^n
\sum_{\lambda\vdash 2n-1,
\sum^{\infty}_{l=1} i_{\lambda}(l) = n}
c^{(n)}_{\lambda}\prod^{\infty}_{l=1} m_l^{i_{\lambda}(l)},
\label{eq:Q_z_final}
\end{eqnarray}
where
\begin{equation}
c^{(n)}_{\lambda} = \frac{(n-1)!}{\prod^{\infty}_{l=1}i_{\lambda}(l)!}.
\end{equation}
Accordingly,
\begin{equation}
q_n = \sum_{\lambda\vdash 2n-1,
\sum^{\infty}_{l=1} i_{\lambda}(l) = n}
c^{(n)}_{\lambda}\prod^{\infty}_{l=1} m_l^{i_{\lambda}(l)}
\label{eq:q_n_final}
\end{equation}
when $n\ge 1$, and $q_0 = 0$.
 
What is necessary for deriving $R(z)$ is just to replace $g(w)=w$ with 
$g(w)=1/(1-w)$ when applying \EQ(\ref{eq:Lagrange}). Using
\EQ(\ref{eq:M_power_n}), we obtain
the annealed approximation form of $R(z)$:
\begin{eqnarray}
\hspace*{-0.8cm}
R(z) &=& \frac{1}{1-Q(z)}\nonumber\\
&=& g(w)\nonumber\\
&=& 1 + 
\sum^{\infty}_{n=1}\frac{z^n}{n!}
\left[ \frac{d^{n-1}}{du^{n-1}} 
\left( \frac{-1}{(1-u)^2}
\left(\frac{M(u)}{u}\right)^n\right) \right]_{u=0}\nonumber\\
&=& 1 + 
\sum^{\infty}_{n=1}\frac{z^n}{n!}
\sum^{n-1}_{n^{\prime\prime}=0}
\left[ \frac{(-1)^{n-n^{\prime\prime}}(n-n^{\prime\prime})!}{(1-u)^{n-n^{\prime\prime}+1}}
\frac{d^{n^{\prime\prime}}}{du^{n^{\prime\prime}}} 
\left(\frac{M(u)}{u}\right)^n \right]_{u=0}\nonumber\\
&=& 1 + 
\sum^{\infty}_{n=1}\frac{z^n}{n!}
\sum^{n-1}_{n^{\prime\prime}=0}
(-1)^{n-n^{\prime\prime}}(n-n^{\prime\prime})!\nonumber\\
&&\times
\left[ \sum^{\infty}_{n^{\prime}=n+n^{\prime\prime}}
\frac{(n^{\prime}-n)!}{(n^{\prime}-n-n^{\prime\prime})!}
u^{n^{\prime}-n-n^{\prime\prime}}
\sum_{\lambda\vdash n^{\prime},
\sum^{\infty}_{l=1} i_{\lambda}(l) = n}
\frac{n!}{\prod^{\infty}_{l=1}i_{\lambda}(l)!}
\prod^{\infty}_{l=1} m_l^{i_{\lambda}(l)}
\right]_{u=0}\nonumber\\
&=& 1 + 
\sum^{\infty}_{n=1}\frac{z^n}{(n-1)!}
\sum^{n-1}_{n^{\prime\prime}=0}
(-1)^{n-n^{\prime\prime}}(n-n^{\prime\prime})!\; n^{\prime\prime}!
\sum_{\lambda\vdash n+n^{\prime\prime},
\sum^{\infty}_{l=1} i_{\lambda}(l) = n}
c^{(n)}_{\lambda} \prod^{\infty}_{l=1} m_l^{i_{\lambda}(l)},
\label{eq:R_Lagrange}
\end{eqnarray}
which results in
\begin{equation}
r_n = \frac{1}{(n-1)!}
\sum^{n-1}_{n^{\prime\prime}=0}
(-1)^{n-n^{\prime\prime}}(n-n^{\prime\prime})!\; n^{\prime\prime}!
\sum_{\lambda\vdash n+n^{\prime\prime},
\sum^{\infty}_{l=1} i_{\lambda}(l) = n}
c^{(n)}_{\lambda} \prod^{\infty}_{l=1} m_l^{i_{\lambda}(l)},
\end{equation}
when $n\ge 1$, and $r_0 = 1$.

\section{Examples}\label{sec:examples}

The analytical methods developed in
\SEC\ref{sec:return} can be broadly applied since the only assumptions
that we have made on $\{p_k \}$ are
$p_0=0$ and that the average vertex
degree is not so small. In this section, we apply our
theoretical estimates
to random walk on some classes of graphs
that are often relevant in real-world situations and also of
theoretical interest.

\subsection{Cayley trees}

Let us first consider the Cayley trees \cite{Albert02}
in which each vertex has exactly $d$
vertices. Substituting $p_k = \delta_{k,d}$ into \EQ(\ref{eq:Q}) yields
\begin{equation}
Q(z) = \frac{1}{d-(d-1)Q(z)}.
\label{eq:Q_cayley1}
\end{equation}
Although \EQ(\ref{eq:Q_cayley1}) has two different solutions of $Q(z)$,
the one satisfying $Q(1)=1$ is excluded
because of the transient nature of the random walk on the Cayley trees 
\cite{Spitzer,Lyons90_Lyons95,Grimmettbook}.
Then \EQ(\ref{eq:Q_cayley1}) is led to
\begin{equation}
Q(z) = \frac{d-\sqrt{d^2-4(d-1)z}}{2(d-1)}.
\label{eq:Q_cayley2}
\end{equation}
In this case, $Q(z)$ is related to the generating function $S(z)$ 
of Catalan numbers $D_n \equiv {}_{2n}C_{n} / (n+1)$ \cite{Stanleybook} 
as follows:
\begin{equation}
S(z) = \frac{1-\sqrt{1-4z}}{2z} 
= \frac{d-1}{dz}Q\left(\frac{d^2}{d-1}z\right).
\label{eq:S_Q}
\end{equation}
Accordingly, we obtain
\begin{equation}
q_n = \frac{(d-1)^{n-1}}{d^{2n-1}}D_{n-1}.
\label{eq:q_n_cayley1}
\end{equation}
On the other hand, applying $m_l = (d-1)^{l-1} / d^l$ to
\EQ(\ref{eq:q_n_final}) results in
\begin{eqnarray}
q_n &=& \sum_{\lambda \vdash 2n-1, \sum^{n}_{l=1}i_{\lambda}(l) = n}
c^{(n)}_{\lambda} \frac{(d-1)^{\sum^{n}_{l=1}(l-1) i_{\lambda}(l) }}
{d^{\sum^n_{l=1}l i_{\lambda}(l)}}\nonumber\\
&=& \frac{(d-1)^{n-1}}{d^{2n-1}}
\sum_{\lambda \vdash 2n-1, \sum^{n}_{l=1}i_{\lambda}(l) = n}
c^{(n)}_{\lambda}.
\label{eq:q_n_cayley2}
\end{eqnarray}
Combining
\EQS(\ref{eq:q_n_cayley1}) and (\ref{eq:q_n_cayley2}) provides a 
useful by-product:
\begin{equation}
\sum_{\lambda \vdash 2n-1, \sum^{n}_{l=1}i_{\lambda}(l) = n}
c^{(n)}_{\lambda} = D_{n-1},
\end{equation}
which states that
the sum of the coefficients in the moment expansion of
$q_n$ [see \EQ(\ref{eq:q_n_final})] is always equal to $D_{n-1}$
without regard to the distribution $\{p_k\}$.

Similarly, \EQ(\ref{eq:R_Lagrange}) becomes
\begin{equation}
R(z) = \frac{2-d+\sqrt{d^2-4(d-1)z}}{2(1-z)}.
\label{eq:R_cayley}
\end{equation}
Then, it follows that
\begin{equation}
\left(1-\frac{d^2}{d-1}z\right) R\left( \frac{d^2}{d-1}z \right)
= 1- d z S(z)
\end{equation}
and
\begin{equation}
r_n = 1 - \sum^{n-1}_{n^{\prime}=0}
\left(\frac{(d-1)^{n^{\prime}-1}}{d^{2n^{\prime}-1}}\right) D_{n^{\prime}}.
\end{equation}
Owing to the entire homogeneity of the Cayley trees,
\EQS(\ref{eq:Q_cayley2}) and (\ref{eq:R_cayley}) are exact in this case
and agree with the theoretical results obtained by identifying 
random walk on the Cayley trees with unbiased random walk on
${\bf Z}$ \cite{Spitzer,Grimmettbook}.

\subsection{Erd\"{o}s-R\'{e}nyi random graph}\label{sub:er}

The Erd\"{o}s-R\'{e}nyi (ER) random graph is generated by independently
assigning an edge with probability $p$ between any possible pairs
of vertices
\cite{Erdos,Albert02,Newman}.  If the number of vertices $N$ scales so
that $\lambda \equiv N p$ converges in the limit $N\to\infty$, the
vertex degree is distributed as specified by the Poiss\'{o}n
distribution, namely,
\begin{equation}
p_k = \frac{\lambda^k}{k!} {\rm e}^{-\lambda}.
\end{equation}

\begin{figure}
\begin{center}
\includegraphics[height=2.25in,width=3.25in]{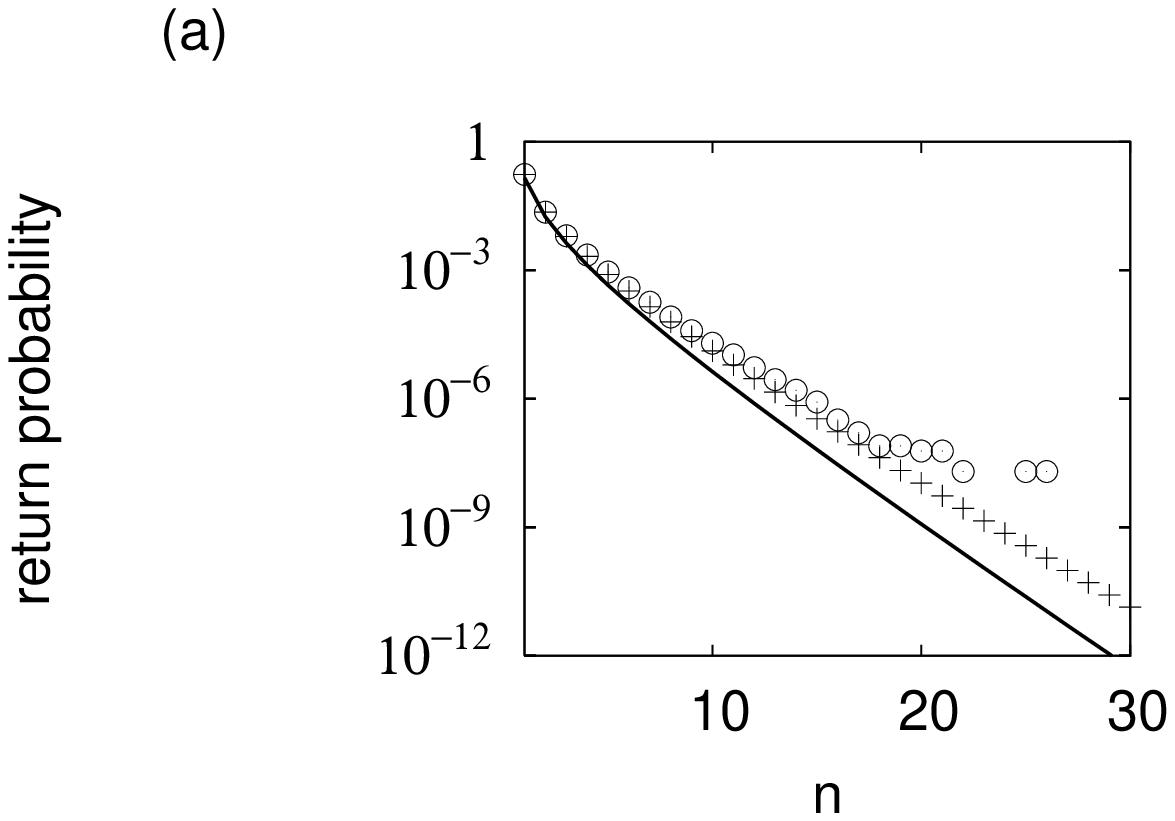}
\includegraphics[height=2.25in,width=3.25in]{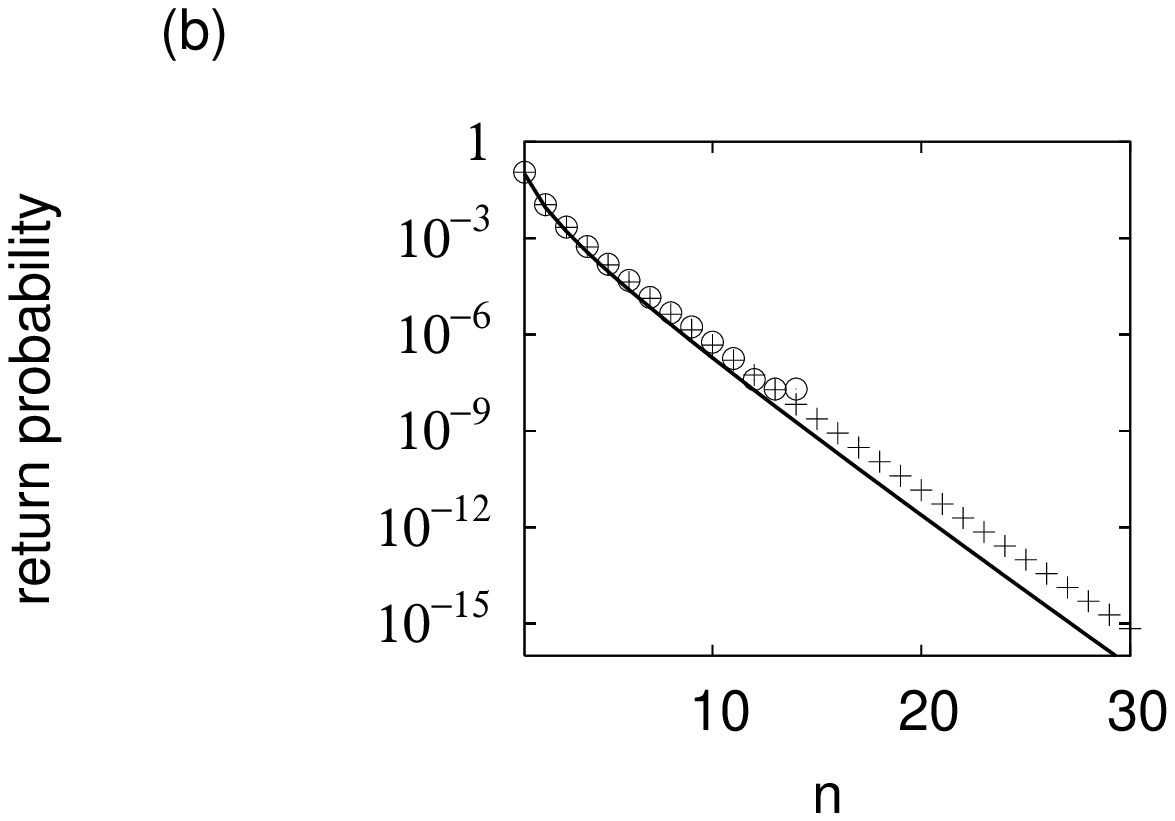}
\caption{}
\label{fig:er}
\end{center}
\end{figure}

Numerically calculated distributions of the first return time are
indicated by circles
in \FIGS\ref{fig:er}(a) and \ref{fig:er}(b) for
$\lambda = 7$ and $\lambda = 10$, respectively.
The return probability decreases exponentially
in $n$ analogous to the case of the Cayley trees indicated by
solid lines in \FIG\ref{fig:er}
\cite{Spitzer,Schinazibook,Grimmettbook}. Then many sample points 
are required for reliable estimation of the return time probability,
for which reason we construct the probability distributions based on
$5\times 10^7$ runs. A new random graph is created in each run.

Distributions predicted by the theory in \SEC\ref{sec:return} are
indicated by crosses in \FIG\ref{fig:er}.  The theoretical estimates agree
with the numerical results better when $\lambda=10$. This is
because our method works better for networks
with a larger mean vertex degree, which is equal to $\lambda$.
However, 
the error is bearable in both cases 
for sufficiently small $n$ 
for which the numerical distributions
are calculated based on enough sample points.
In other words, the minimum positive probability
obtained by the simulations is $1/(5\times 10^7) = 2\times
10^{-8}$, and the numerically 
estimated probabilities are not reliable around this
value where statistical fluctuation counts. Related to this remark,
\FIG\ref{fig:er} shows that the numerical results are actually
available just
up to small values of $n$, that is, $n\le 17$ for $\lambda=7$ and
$n\le 12$ for $\lambda=10$. As noted before, this is due to the
exponential decay in the return time distribution.  Furthermore, the
decay is faster for a larger mean vertex degree, or a larger $\lambda$,
which more severely constrains the practical 
upper limit of $n$ for which the
distribution is obtained. Compared with the cumbersome 
brute-force method, our method needs
only calculation of partition of integers, 
which are much more numerically feasible.

Figure~\ref{fig:er} also shows, both for $\lambda=7$ and $\lambda=10$,
that the decay of the first return time probability is slower for the
ER random graphs than for the Cayley trees
with the same mean vertex degree.
This is presumably because of the dispersion of vertex degree in the
ER random graph, as we
discuss in \SEC\ref{sec:conclusions}.

\subsection{Scale-free networks}

\begin{figure}
\begin{center}
\includegraphics[height=2.25in,width=3.25in]{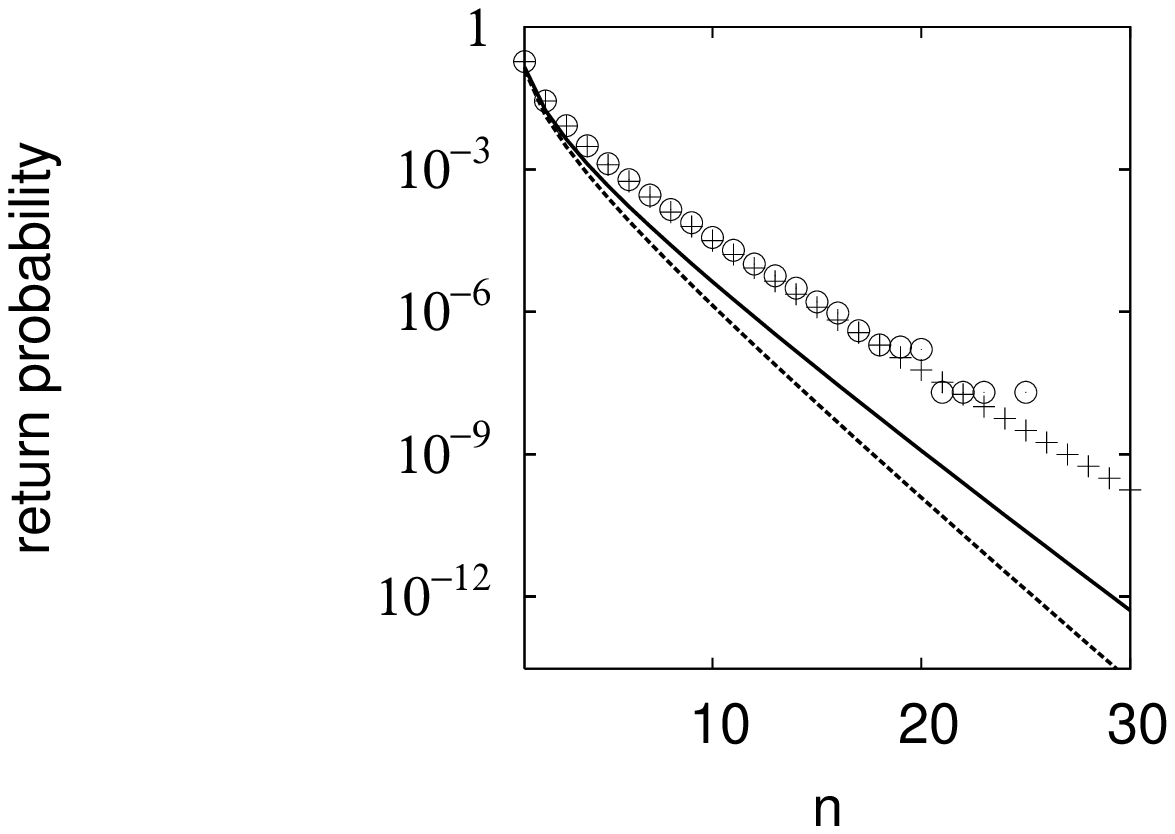}
\caption{}
\label{fig:sf}
\end{center}
\end{figure}

The vertex degrees of real networks often have power-law
distributions. Barab\'{a}si and co-workers presented a network
growth model with preferential attachment to generate such a graph
\cite{SW-SF,Albert02}. In their scale-free networks, the vertex degree
has a lower cutoff $m$, and the degree distribution is represented by
$p_k = {\cal N} k^{-3}$ ($k\ge m$) and $p_k=0$ ($k<m$), where ${\cal
N}$ is the normalization constant. The first return time
probabilities of random walk on scale-free random graphs with
$m=4$ are shown in \FIG\ref{fig:sf}, suggesting that the theory (crosses) again
predicts the numerical results (circles) in a satisfactory manner. In
this case, the mean vertex degree is numerically calculated to be
7.09.  Accordingly, the results for the Cayley trees with $d=7$ (solid
lines) and $d=8$ (dotted lines) are also shown in \FIG\ref{fig:sf} for
comparison.  The probability of the first return time decays slower
for the scale-free networks, as has also been the case for the ER
random graphs. Moreover, comparison of \FIGS\ref{fig:er} and
\ref{fig:sf} reveals that the discrepancy from the regular case, which
is probably caused by the heterogeneous vertex degree, is larger for
the scale-free networks.  This is
presumably because the vertex degree is more heterogeneous in the
scale-free networks than in the ER random graphs.

Random walk on other related graphs,
such as ones whose degree distributions have
power laws without the lower cutoff, power laws with exponential higher cutoff, or
simple exponential decay \cite{Callaway00,Newman}, can be analyzed
similarly. The only caveat is that the theory is likely
to fail when the vertex degree is fairly
small on average. Let us also mention that there is little hope for
obtaining more tractable analytical expressions for $Q(z)$ and $R(z)$ 
even in simpler scale-free cases, because the
polylogarithm functions, which can be estimated only numerically
\cite{Newman}, appear in the calculation of $m_l$.

\section{Conclusions}\label{sec:conclusions}

In this paper, we have derived for generalized random networks the
analytic expressions for the probability distributions of first
and general return times. Our methods correctly predict the
numerical results as far as the mean vertex degree is not extremely
small.  They are also useful in saving the computation time and hence
obtaining return time probabilities on a much longer time scale than
with straightforward simulations. This merit stems from the fact
that the algorithm for calculating partition of integers is easily
implemented \cite{Stanleybook}, whereas brute-force methods require
billions of runs to obtain the distributions and the asymptotics, particularly
in the case of exponentially decaying tails.

We have also found that heterogeneous graphs such as
the ER random graphs and the scale-free networks yield slower
decay of return time probabilities than the Cayley trees with 
the corresponding vertex degrees. The decay rate is closely linked
to critical phenomena and phase transitions of both static
\cite{Lyons90_Lyons95} and dynamical
\cite{Liggettbook99,Schinazibook,Pemantle01,Durrettbook} particle
systems. In social contexts, information and diseases are actually
suggested to propagate in a manner different from as we imagine by the
analogy of regular graphs such as the Cayley trees and regular
lattices. For example, percolation is more likely to occur in networks with
heterogeneous vertex degrees \cite{Newman}. Also for
dynamical processes such as contact
processes \cite{Liggettbook99,Schinazibook,Eguiluz_Pastor_Masuda_SARS}
and voter models \cite{Liggettbook99,Durrettbook},
occurrence of global orders such as
epidemics or unanimity has the same
tendency.  Mathematically, the problem of the global orders emerging in these
dynamics can be associated with that of the dual or related
processes. For example, if simple and branching random walks
(resp. coalescing random walks) are more likely to return to the
origin, the critical value for phase transition becomes smaller, and the
probability of a global epidemics or
unanimity becomes larger in contact processes
\cite{Liggettbook99,Schinazibook} (resp. voter models
\cite{Liggettbook99,Durrettbook}).  Accordingly, the asymptotic behavior
of random walk reported in \SEC\ref{sec:examples} suggests that global
orders are more likely consequences in networks with heterogeneous
vertex degrees such as scale-free and ER random networks. This
evidence substrates the results for the contact processes in
epidemic contexts \cite{Eguiluz_Pastor_Masuda_SARS} and poses a
dynamical version of the exact results on percolation \cite{Newman}.

As for exact asymptotic behavior, questions about
the Cayley trees with vertex degree $d$
is translated into ones about the unbalanced random walk on
${\bf Z}$, the analysis of which easily resulting in $r_n\propto n^{-3/2}
\left(2\sqrt{d-1} / d\right)^{2n}$ \cite{Schinazibook}.  To
illuminate the asymptotic behavior of $q_n$ and $r_n$ in the case of
generalized random walks is an important subject of future work.

\begin{acknowledgements}
We thank H. Kesten and G. F. Lawler for helpful comments.
This study is supported by the Grant-in-Aid for Scientific Research
(JSPS) and the Grant-in-Aid for Scientific Research (B)
(Grant No. 12440024) of the Japan Society of the Promotion of Science.
\end{acknowledgements}


\bigskip
\bigskip

Figure captions

\bigskip

Figure 1: Schematic diagram showing random walk on a realization of 
generalized random graph. Integers denotes the time of random walk.

\bigskip

Figure 2: Probability distributions of the first return times of the
random walk on the ER random graph with (a)
$\lambda = 7$ and (b) $\lambda = 10$. Numerical and theoretical
results are indicated by circles and crosses, respectively. The results
for the Cayley trees with the same
mean vertex degrees, namely, (a) $d=7$ and (b) $d=10$,
are indicated by solid lines.

\bigskip

Figure 3: Probability distributions of the first return times in the
case of scale-free networks with $m=4$. Numerical and theoretical
results are indicated by circles and crosses, respectively.
The results for the Cayley trees with 
$d=7$ (solid lines) and $d=8$ (dashed lines) are also shown.


\bigskip
\bigskip

\end{document}